\documentclass[preprint,aps]{revtex4}
\usepackage{amsfonts}
\usepackage{graphics}
\usepackage{graphicx}
\usepackage{epsf}
\newcommand{\bea}{\begin{eqnarray}}
\newcommand{\eea}{\end{eqnarray}}
\newcommand{\pa}{\partial}

\begin{document}
\title{On the duality in four-dimensional Lorentz-breaking field theories}

\author{M. S. Guimaraes$^{a}$, J. R. Nascimento$^{b}$, A. Yu. Petrov$^{b}$, C. Wotzasek$^{c}$}
\email{msguimaraes@uerj.br, jroberto@fisica.ufpb.br, petrov@fisica.ufpb.br, clovis@if.ufrj.br}
\affiliation{$^a$ UERJ - Universidade do Estado do Rio de Janeiro, Instituto de F\'isica - Departamento de F\'isica Te\'orica, Rua S\~ao Francisco Xavier 524, 20550-013 Maracan\~a,
Rio de Janeiro, Brasil\\$^b$ Departamento de F\'\i sica, Universidade Federal da Para\'\i ba,
Caixa Postal 5008, 58051-970 Jo\~ao Pessoa, Para\'\i ba, Brazil\\$^c$Instituto de F\'\i sica, Universidade Federal de Rio de Janeiro,
Caixa Postal 21945, Rio de Janeiro, Brazil}
%\date{\today}

\begin{abstract}
We consider new issues of duality in four-dimensional Lorentz-breaking field theories. In particular, we demonstrate that the arising of the aether-like Lorentz-breaking term is necessary in order for the 4D models to display the duality analog between the MCS and self-dual models in 3D. We further study the dispersion relations in both theories and discuss the physical contents of the models involved in this new dualilty.
\end{abstract}

\maketitle
\section{Introduction}
The duality between two different field theory models is an important concept allowing for mutual mapping of theories possessing essentially different actions. In three dimensions, the duality was initially established for the free self-dual and Maxwell-Chern-Simons theories \cite{DJ}.  Later the duality was observed for a wide class of extensions of these theories, including coupling of the self-dual and Maxwell-Chern-Simons fields to the scalar \cite{imm1} and spinor matter \cite{imm1,Mal}, nonlinear generalizations of these models \cite{N2}, their noncommutative extension \cite{N3,N4}, including of the Lorentz symmetry breaking \cite{N5}, and supersymmetric extension constructed on the base of the superfield formalism \cite{N4,imm2}. Beside of these generalizations of the self-dual and Maxwell-Chern-Simons theories, the duality was also established for the higher-derivative models \cite{N7} and higher-rank tensor field models \cite{N8}.

In all the examples above, the 3D duality involved the presence of the Chern-Simons term.
%All these results were obtained for the three-dimensional field theories. Moreover, up to now the duality was treated as a purely three-dimensional phenomenon since the Chern-Simons term which is a key ingredient of the self-dual and Maxwell-Chern-Simons theories is essentially three-dimensional.
However, the possibility of Lorentz symmetry breaking opens new perspectives for implementation of this kind of duality in four-dimensional field theories. The key idea consists in using of the Lorentz-breaking Carroll-Field-Jackiw (CFJ) term \cite{CFJ} which is a Chern-Simons-like term defined in four dimensions and possessing the gauge symmetry similarly to the Chern-Simons term. In this paper we establish duality between four-dimensional analogs of the self-dual and Maxwell-Chern-Simons theories where the Carroll-Field-Jackiw term is used instead of the Chern-Simons term.

The paper is organized as follows. The Section 2 is devoted to the development of the gauge embedding method for the Lorentz-violating self-dual theory which leads to an aether-like generalized electrodynamics. In Section 3 the propagators and dispersion relations in the new theory are analyzed. The Section 4 is devoted to detailed study and comparison of the massive modes in both theories, and in the section 5 the duality is confirmed by use of the inverse mapping of the new theory to the Lorentz-breaking self-dual theory. In the Summary, the results are discussed.

\section{Gauge embedding for the Lorentz-breaking self-dual theory}
Let us consider the gauge embedding of the four-dimensional Lorentz-breaking self-dual theory. This model has the Lagrangian
\bea
\label{sd4}
L_{SD}=\frac{m^2}{2}f^af_a-\frac{1}{2}\epsilon^{abcd}b_af_b\pa_c f_d+f^aj_a,
\eea
where $f_a$ is the self-dual vector field, $b_a$ is a mass dimensional constant vector breaking the Lorentz symmetry, and $j_a=\bar{\psi}\gamma_a\psi$ is a current formed by spinor field $\psi$. This Lagrangian is a natural four-dimensional generalization of the three-dimensional self-dual action \cite{imm1}, in which the Chern-Simons term is replaced by its four-dimensional Lorentz-breaking analog \cite{JK}.

It is interesting, at this junction, to check that this 4-dimensional extension indeed share the self-duality property of its 3D analog.
The equations of motion for this theory look like
\bea
\label{eqmot}
m^2f^a = - \epsilon^{abcd}b_b\pa_cf_d,
\eea
which plays the role of self-duality condition and also shows that the self-dual field satisfy the conditions
\bea
\label{propert}
\pa_a f^a = b_a f^a = 0.
\eea
If we apply the self duality condition again on the right hand side of (\ref{eqmot}) we obtain
\bea
m^4f_a=-[b_a(b\cdot\pa)(\pa\cdot f)-b^cb_a\Box f_c+b^c(b\cdot \pa)\pa_a f_c-b^c(b\cdot\pa)\pa_cf_a+b^2\Box f_a-b^2\pa_a\pa_c f^c]\nonumber
\eea
which, after using conditions (\ref{propert}) diagonalizes to become
\bea
 [(b\cdot\pa)^2 - b^2\Box - m^4 ]f_a = 0.
\eea

One can find that the Lorentz-breaking Chern-Simons-like term $-\frac{1}{2}\epsilon^{abcd}b_af_b\pa_c f_d$
possesses the natural gauge symmetry
\bea
\label{gt}
\delta f_a=\pa_a\xi,
\eea
with $\xi$ is a gauge parameter. So our aim is to construct the gauge theory on the base of the model (\ref{sd4}). To do this we follow the gauge embedding formalism \cite{imm1,Ilha:2001he}.
We start introducing the Euler vector which is the kernel of the equation of motion:
\bea
\label{euler}
K^a=m^2f^a+\epsilon^{abcd}b_b\pa_cf_d+j^a.
\eea
Under the gauge transformations (\ref{gt}) the variation of this vector is $\delta K_a=m^2\pa_a\xi$.
The first-order iterated Lagrangian is introduced as
\bea
L^{(1)}=L_{SD}-\Lambda^aK_a,
\eea
where $\Lambda^a$ is a Lagrange multiplier. Since the variation of the initial Lagrangian is $\delta L=\frac{\delta L_{SD}}{\delta f_a}\delta f_a=K^a\pa_a\xi$, one can suggest the transformation for $\Lambda_a$ to be $\delta\Lambda_a=\pa_a\xi$ and write down the variation of $L^{(1)}$ as
\bea
\delta L^{(1)}=-\Lambda^a\delta K_a=-m^2\Lambda^a\delta\Lambda_a.
\eea
Thus, we have the second-order iterated Lagrangian of the form
\bea
L^{(2)}=L^{(1)}+\Delta L^{(2)},
\eea
whose variation is equal to zero if and only if $\delta \Delta L^{(2)}=m^2\Lambda^a\delta\Lambda_a$, or, as is the same,
\bea
\Delta L^{(2)}=\frac{1}{2}\Lambda^a\Lambda_a.
\eea
Thus, the whole second-order iterated Lagrangian is
\bea
L^{(2)}=L_{SD}-\Lambda^aK_a+\frac{1}{2}\Lambda^a\Lambda_a.
\eea
We can eliminate the Lagrange multiplier $\Lambda_a$ via its equations of motion
\bea
m^2\Lambda_a-K_a=0,
\eea
which gives the second-order iterated Lagrangian of the form
\bea
L^{(2)}=L_{SD} -\frac{1}{2m^2}K^aK_a.
\eea
Substituting here the Euler vector $K_a$ from (\ref{euler}), we arrive at
\bea
L^{(2)}=\frac{1}{2}\epsilon^{abcd}b_af_b\pa_cf_d
+\frac{1}{2m^2}j^aj_a+\frac{1}{m^2}\epsilon_{abcd}b_aj_b\pa_cf_d-
\frac{1}{2m^2}\epsilon^{abcd}\epsilon_{ab'c'd'}b^{b'}b_b\pa^{c'}f^{d'}\pa_cf_d.
\eea
Multiplying the Levi-Civita symbols, for the signature $(+---)$, introducing the stress tensor $F_{ab}=\pa_af_b-\pa_bf_a$, and relabelling $f_a\to A_a$, we get the following Lagrangian $L_{ED}$ for the new generalized Lorentz-violating electrodynamics:

\bea
\label{new}
L_{ED}&=&\frac{1}{2}\epsilon^{abcd}b_aA_b\pa_cA_d+\frac{1}{2m^2}j^aj_a+\frac{1}{2m^2}\epsilon^{abcd}b_aj_b F_{cd}+\frac{b^2}{4m^2}F^{ab}F_{ab}-\nonumber\\&-&\frac{1}{2m^2}b^cb_bF^{bd}F_{cd}.
\eea

\noindent We note that this theory becomes trivial in the Lorentz-covariant limit $b^a=0$, which is very natural since the initial self-dual  action (\ref{sd4}) possesses nontrivial dynamics only for non-zero $b^a$.  This action is composed by the following terms: the CFJ term, the Thirring-like current-current interaction, a magnetic non-minimal coupling, the Maxwell term and finally a new term described below.

The arising of the new, CPT-even Lorentz-breaking term $-\frac{1}{2m^2}b^cb_bF^{bd}F_{cd}$ is a nontrivial result. This is a perfect example of the aether-like terms which probably are very important in the context of the presence of the compact extra dimensions \cite{Carroll}, note however that in our case this term arises already in four-dimensional space-time. Further we will refer to this term as to the aether-like term, and to this theory as to the electrodynamics with aether-like term. We note that some implications of presence of the aether-like term were discussed in \cite{Andrianov} in the phenomenological context; however, the complete action studied in \cite{Andrianov} was not gauge invariant. Also, the aether-like terms, in particular, in the case of the electrodynamics, were shown in \cite{ouraether} to arise as perturbative corrections in different dimensions.

\section{Propagator and dispersion relations in the electrodynamics with aether-like term}
Let us study some properties of the theory described by the Lagrangian (\ref{new}), involving the aether-like term.
First, we can find the equations of motion for the Lagrangian $L_{ED}$:
\bea
-\frac{1}{2}\epsilon_{abcd}b^bF^{cd}-\frac{1}{m^2}b^b\epsilon_{abcd}\pa^c j^d+\frac{1}{m^2}b^bb^c\pa_bF_{ca}-\frac{b^2}{m^2}\pa^bF_{ba}=0.
\eea
We find that if $|b|\simeq m$, and the mass $m$ is high enough, the Maxwell term together with new Lorentz-breaking term would be the dominant ones.
Second, we can find a propagator for this theory. To do this, we can, following \cite{Jack}, split the vector field into longitudinal and transversal parts:
\bea
A_a=\bar{A}_a+\pa_a \lambda,
\eea
with $\pa_a\bar{A}^a=0$. As a result, the Lagrangian (\ref{new}), in the case of zero currents, takes the form
\bea
L_{ED}=\frac{1}{2}\bar{A}_a\left(-\frac{b^2}{m^2}\eta^{ab}\Box+\frac{1}{m^2}(b\cdot \pa)^2\eta^{ab}+\frac{1}{m^2}b^ab^b\Box+\epsilon^{kalb}b_k\pa_l
\right)\bar{A}_b.
\eea
We note that the $\lambda$ dependent (longitudinal) part of the Lagrangian totally vanishes. It remains to find the inverse operator to
\bea
\Delta^{ab}=-\frac{b^2}{m^2}\eta^{ab}\Box+\frac{1}{m^2}(b\cdot \pa)^2\eta^{ab}+\frac{1}{m^2}b^ab^b\Box+\epsilon^{kalb}b_k\pa_l.
\eea
Straightforward calculations show this inverse operator $G_{bc}$ (such as $\Delta^{ab}G_{bc}=\delta^a_c$) to be
\bea
G_{bc}=X_1\eta_{bc}+X_2b_bb_c+X_3\epsilon_{mbnc}b^m\partial^n+X_4\pa_b\pa_c+X_5b_b\pa_c+X_6b_c\pa_b,
\eea
with
\bea
X_1&=&\frac{A_1}{A^2_1-(b^2\Box-(b\cdot\pa)^2)};\nonumber\\
X_3&=&-\frac{1}{A^2_1-(b^2\Box-(b\cdot\pa)^2)};\nonumber\\
X_4&=&-\frac{b^2}{A_1[A^2_1-(b^2\Box-(b\cdot\pa)^2)]};\nonumber\\
X_6&=&\frac{(b\cdot\pa)}{A_1[A^2_1-(b^2\Box-(b\cdot\pa)^2)]}.
\eea
Here
\bea
A_1&=&\frac{1}{m^2}[(b\cdot\pa)^2-b^2\Box],\quad\, A_2=\frac{\Box}{m^2}.
\eea
The $X_2$ and $X_5$ have more complicated structure:
\bea
X_2&=&\frac{-A_2X_6(b\cdot\pa)-A_2X_1+A_3X_3\Box}{A_1+A_2b^2};\nonumber\\
X_5&=&\frac{-A_2X_4-A_3X_3(b\cdot\pa)}{A_1+A_2b^2}.
\eea
It is easy to see that these expressions have the common denominator $A^2_1-(b^2\Box-(b\cdot\pa)^2)$. So, these expressions, after Fourier transform, correspond to the following dispersion relations:
\bea
\label{disprel}
[b^2k^2-(b\cdot k)^2][b^2k^2-(b\cdot k)^2+m^4]=0.
\eea
Thus, there are two possibilities:\\
\bea
\label{disprel1}
&&{\rm(i)} \,\,\, b^2k^2-(b\cdot k)^2=0;\nonumber\\
&&{\rm (ii)}\, b^2k^2-(b\cdot k)^2+m^4=0.
\eea
One can see that the time-like $b^{\mu}=(b_{0},\vec{0})$ in the case (i) does not correspond to physically consistent dispersion relations (the space momentum would be zero with no relation to energy), and in the case (ii) the momentum satisfies the relation
\bea
|\vec{k}|^2=\frac{m^4}{b^2_{0}},
\eea
with the energy is not determined.

As for the space-like case, for example $b^{\mu}=(0,\vec{b})$, in the case (i) it gives the dispersion relations $\omega^{2}= |\vec{k}|^{2}\sin^2\theta$, where we have used the relation: $\vec{b}\cdot\vec{k}=|\vec{b}||\vec{k}|\cos\theta$, with $\theta$ be an angle between $\vec{b}$ and $\vec{k}$,  and in case (ii) the energy satisfies the relation
\bea
\omega^{2}=|\vec{k}|^{2}\sin^{2}\theta+\frac{m^{4}}{|\vec{b}|^{2}}.
\eea
However, we note that the Lorentz symmetry breaking cannot be considered as small correction to the standard dispersion relations. We find that in the case of the plane wave orthogonal to the $\vec{b}$ vector, we have the usual Lorentz-invariant massless dispersion relations $\omega^2=\vec{k}^2$ in the case (i) and usual Lorentz-invariant massive dispersion relation $\omega^2=\vec{k}^2+\frac{m^{4}}{|\vec{b}|^{2}}$ in the case (ii).

\section{Massive modes of the self-dual model}

Now, to confirm the duality, let us study the spectrum of the self-dual model. It is necessary (see for example \cite{N5,Hel}), that the physical sectors of the spectra of dual models should coincide.
Let us discuss the problem of massive modes in the self-dual model, whose Lagrangian is given by (\ref{sd4}).
The equations of motion corresponding to this Lagrangian are
\bea
\label{eqmotion}
f_{\mu}=-\frac{1}{m^2} b^\nu \epsilon_{\mu\nu\rho\lambda} \partial^\rho f^\lambda.
\eea
To find the spectrum of the $A_{\mu}$ field, we recall the following conditions that identically satisfied by the equations of motions (\ref{eqmotion})
\bea
\partial_\mu f^\mu &=&0; \,\,\,\,\,\,\,\,\,\,\,  \text{Lorentz gauge}\label{cond1} \\
b_\mu f^\mu &=&0,        \,\,\,\,\,\,\,\,\,\,\,  \text{axial gauge}\label{cond2}
\eea
and iterate (\ref{eqmotion}) to obtain
\bea
%A_{\mu}&=\frac{1}{m^2} b^\nu \epsilon_{\mu\nu\rho\lambda} \partial^\rho \left[\frac{1}{m^2} b_\theta \epsilon^{\lambda\theta\omega\alpha} \partial_\omega A_\alpha \right] \nn\\
%m^4 A_{\mu}&= \epsilon_{\mu\nu\rho\lambda} \epsilon^{\lambda\theta\omega\alpha} \,   b_\theta \,b^\nu \,\partial^\rho \, \partial_\omega \,A_\alpha  \nn\\
%m^4 A_{\mu}&= \delta^\theta_{[\mu}\delta_\nu^\omega \delta^\alpha_{\rho]}  \,   b_\theta \,b^\nu \,\partial^\rho \, \partial_\omega \,A_\alpha  \\
m^4 f_{\mu}&=& \underbrace{b^\nu b_\mu \partial^\rho \partial_\nu f_\rho} + \overbrace{b^\nu b_\rho \partial^\rho \partial_\mu f_\nu }
             + b^\nu b_\nu \partial^\rho \partial_\rho f_\mu
             - \overbrace{b^\nu b_\mu \partial^\rho \partial_\rho f_\nu} - b^\nu b_\rho \partial^\rho \partial_\nu f_\mu
             - \nonumber\\ &-&\underbrace{b^\nu b_\nu \partial^\rho \partial_\mu f_\rho}.
\eea
Here the marked terms vanish under conditions (\ref{cond1}) (the  underbrace) and (\ref{cond2}) (the overbrace). The remaining term is
\bea
%m^4 A_{\mu}&=& b^\nu b_\nu \partial^\rho \partial_\rho A_\mu - b^\nu b_\rho \partial^\rho \partial_\nu A_\mu, \nn\\
m^4 f_{\mu}&=& b^\nu b_\nu \Box f_\mu - (b^\nu \partial_\nu) (b^\rho \partial_\rho)  f_\mu,
\eea
which we can rearrange as
\bea
[b_\nu b^\nu \Box - (b^\nu \partial_\nu) (b^\rho \partial_\rho) - m^4]f_\mu = 0.
\eea
This expression after the Fourier transform gives exactly the second relation  (\ref{disprel1}) of the electrodynamics with the aether-like term (\ref{new}), that is, just the theory generated via the gauge embedding of the self-dual model (\ref{sd4}). Therefore we can conclude that the duality of these models is confirmed via coincidence of their dispersion relations. We note that the coincidence of physical spectra of both models is also a sufficient condition to confirm the duality \cite{Hel}.

Now, let us develop an adequate treatment of this expression, and hence, of the physical spectrum of both models. To do it, let us recall some properties of the Maxwell model. The equation of motion in the ``Lorentz gauge" looks like
\bea
\Box f_\mu =0 .
\eea
Now let us introduce the plane wave ansatz: $f_\mu = e_\mu \exp({i k_\nu x^\nu})$, with
\bea
-k_\nu k^\nu e_\mu =0  \,\,\,\,\to \,\,\,\,\, k_\nu k^\nu = 0 \,\,\,\,\to \,\,\,\,\, \omega^2 = \vec{k}^2
\eea
The group velocity is $v_g=\frac{d\omega}{d k}=\pm 1$, i.e. exactly the speed of light in a vacuum.

In the Proca theory, the equation of motion is
\bea
(\Box +m^2) f_\mu =0
\eea
which implies in
\bea
(-k_\mu k^\mu +m^2) e_\mu =0  \,\,\,\,\to \,\,\,\,\, k_\nu k^\nu = m^2 \,\,\,\,\to  \,\,\,\,\, \omega^2 = \vec{k}^2+m^2,
\eea
that is, the usual relativistic dispersion relation for the massive case.
The velocity of the particle is
\bea
v_g = \pm \frac{k}{\sqrt{k^2+m^2}}
\eea
This velocity is always less than the speed of light, so we can convince ourselves that the theory is massive.

In our case,
\bea
[-b_\nu b^\nu k_\rho k^\rho + (b^\nu k_\nu)^2 - m^4]e_\mu = 0,
\eea
which yields
\bea
\label{yi}
-b_\nu b^\nu k_\rho k^\rho + (b^\nu k_\nu)^2 = m^4.
\eea
Let us define the four-vectors $k_\mu = (\omega,\vec{k})$ and $b_\mu = (\lambda, \vec{b})$. Also, we introduce a constant $\alpha$ such as $\lambda = \alpha b$ where $b=|\vec{b}|$. It is clear that if $\alpha<1$, the four-vector $b_\mu$ is space-like, if $\alpha=1$, it is light-like, and if  $\alpha>1$ -- time-like. Finally, let $\theta$ be an angle between $\vec{b}$ and $\vec{k}$.
It allows us to rewrite the above expression as

\bea
%-(\lambda^2 - b^2)(\omega^2 - k^2) + (\lambda \omega - bk \cos\theta)^2 &=& m^4 \nn\\
%-(\alpha^2-1)b^2 (\omega^2 - k^2) + (\alpha b\, \omega - bk \cos\theta)^2 &=& m^4 \nn\\
%-(\alpha^2-1) (\omega^2 - k^2) + (\alpha \, \omega - k \cos\theta)^2 &=& (m^2/b)^2 \nn\\
%-(\alpha^2-1) (\omega^2 - k^2) + \alpha^2 \omega^2 + k^2 \cos^2\theta -2\alpha \, \omega k \cos\theta &=& (m^2/b)^2 \nn\\
%\omega^2 +(\alpha^2 -\sin^2\theta) ( k^2)   -2\alpha \, \omega k \cos\theta = (m^2/b)^2 \nn\\
\omega^2   -2\alpha \,  k \cos\theta\, \omega+(\alpha^2 -\sin^2\theta)  k^2  -  (m^2/b)^2 = 0,
\eea
with the solutions of these equations look like
\bea
\label{sols}
%\omega&=&\alpha\,k\cos \theta \pm \sqrt{\alpha^2  k^2  \cos^2 \theta + (m^2/b)^2- k^2(\alpha^2-\sin^2 \theta )} \nn\\
\omega&=&\alpha\,k\cos \theta \pm \sqrt{(1-\alpha^2)  k^2  \sin^2 \theta + (m^2/b)^2},
\eea
so, the group velocity is
\bea
v_g=\alpha \cos \theta \pm \frac{(1-\alpha^2)k \sin^2 \theta}{\sqrt{(1-\alpha^2)  k^2  \sin^2 \theta + (m^2/b)^2}}
\eea

Let us discuss following characteristic situations:

1. $\theta=0$, and the $\vec{k}$ is parallel to $\vec{b}$
\bea
\omega=\alpha\,k\ \pm m^2/b  \,\,\,\,\,\,\,\,\,\,\,\,\,\,\, v_g = \alpha,
\eea
or, in terms of the original variables,
\bea
\omega=(\lambda/b) \,k\ \pm m^2/b  \,\,\,\,\,\,\,\,\,\,\,\,\,\,\, v_g = \lambda /b.
\eea
One can see that the propagation is chiral, i.e. velocity can be either positive or negative, but not two possibilities simultaneously. Other important conclusion is that the photon has a constant velocity, possessing at the same time a non-zero rest mass. If $b<\lambda$ (i.e. $b_\mu$ is time-like), the photon propagates in a superluminal manner. If $b>\lambda$, it propagates with a velocity less than the speed of light. For $b=\lambda$, the photon propagates with an usual speed of light despite it has a mass (this phenomenon also takes place in the two-dimensional Lorentz-breaking scalar field theory model \cite{PP}).

2. Let $\theta=\pi/2$, $\vec{k}$ is orthogonal to $\vec{b}$, and $\alpha<1$:
\bea
\omega^2=(1-\alpha^2)k^2 + (m^2/b)^2   \,\,\,\,\,\,\,\,\,\,\,\,\,\,\, v_g= \pm \frac{(1-\alpha^2)k }{\sqrt{(1-\alpha^2)  k^2  + (m^2/b)^2}}.
\eea
For $\alpha<1$, the maximal possible velocity is $\sqrt{1-\alpha^2}$. For $\alpha=0$, it is an usual behaviour of the massive vector field.

3. $\alpha=0$, only space component.
\bea
\omega^2=  k^2  \sin^2 \theta + (m^2/b)^2 \,\,\,\,\,\,\,\,\,\,\,\,\,\,\, v_g=  \pm \frac{k \sin^2 \theta}{\sqrt{k^2  \sin^2 \theta + (m^2/b)^2}},
\eea
if $\theta=0$ there is no propagation. If $\theta=\pi/2$, the propagation is the same as in the massive vector field case. For all values of $\theta$, the velocity of propagation is always less than the speed of light.

4. $\alpha \to \infty$, only temporal component.
The dynamics in this case can be read off from (\ref{yi}), whereas the expression (\ref{sols}) is much less convenient in this case. It follows from (\ref{yi}) that $k=\pm \frac{m^2}{\lambda}$, whereas $\omega$ totally decouples and does not depend on $k$, so $v_g=0$.

5. $\alpha=1$, $b_\mu$ is light-like.
\bea
\omega=k\cos \theta \pm m^2/b  \,\,\,\,\,\,\,\,\,\,\,\,\,\,\, v_g=\cos \theta.
\eea
Again, the propagation is chiral, i.e., the velocity is either positive or negative, but not two cases simultaneously. Also, the photon has a constant velocity and the non-zero rest mass. The velocity is always less than the speed of light except of the case $\theta=0$ where the photon propagates with the speed of light despite it has a mass.

\section{Inverse mapping of the aether-like term}

In this section we show by a direct calculation that the electrodynamics with the aether-like term is the dual formulation of the Lorentz-breaking self-dual theory.

First, it is interesting to construct the dual of the simple aether-like model (without the CS-like term) whose Lagrangian is
\bea
L=-\frac{1}{4}F^{ab}K_{abcd}F^{cd}=-\frac{1}{2}(\epsilon_{abcd}b^b\partial^c A^d)^2 \label{kost}
\eea
with $b_a$ a vector implementing the Lorentz symmetry breaking and
\bea
K_{abcd}=g_{ac}g_{bd}b^2+2b_bb_c g_{ad}.
\eea
This Lagrangian can be rewritten in a reduced order form, with the introduction of an auxiliary field $\pi_a$:
\bea
\label{red}
L=\pi_a\epsilon_{abcd}b^b\partial^c A^d+\frac{1}{2}\pi^a\pi_a.
\eea
We can eliminate $A^a$ via its equations of motion:
\bea
\epsilon^{abcd}b_b\partial_c\pi_d=0,
\eea
which yields $\pi_a=\partial_a\phi+b_a\psi$. Substituting this expression into the Lagrangian (\ref{red}) we find
\bea
L=\frac{1}{2}(\partial_a\phi)^2+\psi(b\cdot\partial)\phi+\frac{b^2}{2}\psi^2.
\eea
We further eliminate the $\psi$ field via its equation of motion:
\bea
\psi=-\frac{(b\cdot\partial)\phi}{b^2},
\eea
thus, our Lagrangian takes the form
\bea
L=-\frac{1}{2}\phi(\Box-\frac{b^ab^b}{b^2}\partial_a\partial_b)\phi.
\eea
It is well known that the usual (Lorentz symmetric) $4D$ abelian gauge theory is self-dual, that is, its dual formulation is another $4D$ abelian gauge theory of the same form. It is thus interesting to note that even though (\ref{kost}) is a (Lorentz violating) $4D$ gauge theory, its dual formulation is in terms of a scalar field. This of course can be immediately traced to the presence of the massive Lorentz violating vector parameter $b^a$ and its role in the above derivation, effectively reducing the rank of the auxiliary field $\pi^a$.

Remembering that the (Lorentz symmetric) $3D$ Maxwell gauge theory is dual to a $3D$ scalar theory, it seems that the presence of the Lorentz violating parameter $b^a$ gives a $3D$ flavor to these $4D$ gauge theories. In fact we will now see that this analogy carries on to another famous $3D$ duality. Consider the complete theory involving both aether-like term and the Carroll-Field-Jackiw term. However to check the duality it is enough to consider the currrent-free case,
\bea
L_{KCFJ}=-\frac{1}{2}(\epsilon_{abcd}b^b\partial^c A^d)^2-A_a\epsilon^{abcd}b_b\partial_c A_d,
\eea
whose equivalent reduced form is
\bea
L=\pi^a\epsilon_{abcd}b^b\partial^c A^d+\frac{1}{2}\pi^a\pi_a-A_a\epsilon^{abcd}b_b\partial_c A_d.
\eea
We eliminate $A^a$ via its equations of motion:
\bea
\epsilon^{abcd}b_b\partial_c(\pi_d-2A_d)=0,
\eea
whose general solution has the form
\bea
A_a=\frac{1}{2}\pi_a+\partial_a\phi+b_a\psi,
\eea
with $\phi,\psi$ arbitrary scalar fields. Substituting this expression into the Lagrangian above we find that its two last terms vanish, and the final expression looks like
\bea
L=\frac{1}{2}\pi^a\pi_a+\frac{1}{4}\pi_a\epsilon^{abcd}b_b\partial_c \pi_d.
\eea
This is exactly the self-dual Lorentz-breaking action (\ref{sd4}) and is the analog in $4D$ of the well known $3D$ self-dual action \cite{Townsend:1983xs}. Noting that the starting action is, similarly, the $4D$ analog of the $3D$ Maxwell-Chern-Simons action \cite{Deser:1981wh}, the duality we just proved is the exact analog of the famous Maxwell-Chern-Simons / Self-dual duality \cite{Deser:1984kw}. This confirms the duality of the Lorentz-breaking electrodynamics with aether-like term and the self-dual Lorentz-breaking theory.

\section{Summary}

In this work we have established the duality between the four-dimensional Lorentz violating self-dual theory defined by (\ref{sd4}) and the four-dimensional Lorentz violating gauge theory defined by (\ref{new}). This was done by first embedding a gauge symmetry in the non-gauge theory (\ref{sd4}). It turns out that this procedure does not alter the physical properties of the theory, if defined in a topologically trivial space, since all the gauge redundancy is of a topological origin. The equivalence between these two theories was further confirmed by a study of its spectrum and physical properties as displayed in their dispersion relations. We have also provided a direct computation of the duality for the free case.

This result can be understood as a natural generalization of the well known Maxwell-Chern-Simons/Self-dual duality in three-dimensions. In fact, as it happens in $3D$, this is an important result establishing the equivalence between a gauge theory and a non-gauge theory, with the extra gauge redundancy information also of a topological origin.
 
It is important to note that the analogy is far-reaching indeed. The Maxwell-Chern-Simons theory is an effective theory for low energy $QED$ in $3D$, with the Chern-Simons term stemming from radiative corrections originating from the Parity violating fermionic masses. The Chern-Simons term is the first term in a derivative expansion of the fermionic determinant. Similarly, the Lorentz violating gauge theory (\ref{new}) can be understood as an effective theory for the Lorentz violating $QED$ in $4D$. The CPT-odd Carroll-Field-Jackiw term is the first in a derivative expansion of the CPT-odd fermionic determinant. A study of the duality properties of the Carroll-Field-Jackiw theory, that is, without the aether term, was pursued by some of us in \cite{Guimaraes:2006gj}. The present result shows that a much more complete analogy with the $3D$ counterpart can be attained if the aether term is taken into account as well. This is only natural in an effective field theory framework, where the lore is to keep all the terms allowed by the symmetries of the theory up to the given order. The ether term is the next non-trivial Lorentz violating term in the derivative expansion of the fermionic determinant up to second order and so it must be present.

%We have established duality between four-dimensional Lorentz symmetry vioting self-dual theory and a new gauge theory being a  CPT-even Lorentz-breaking modification of the electrodynamics. Thus, we, first time in the literature, established the duality for the case of the four-dimensional field theories defined in (\ref{sd4}).
%
%First, we have shown that, within the gauge embedding procedure, the new, CPT-even Lorentz-breaking term arises, that is, the aether-like term. We confirmed the duality of the new theory with an initial one, that is, the Lorentz-breaking self-dual theory, via inverse mapping procedure. Second, we have studied the dispersion relations and showed that both these theories display very unusual, but coinciding, dispersion relations, which confirms their duality.

{\bf Acknowledgements.} This work was partially supported by Conselho
Nacional de Desenvolvimento Cient\'{\i}fico e Tecnol\'{o}gico (CNPq), Coordena\c{c}\~{a}o de Aperfei\c{c}oamento do Pessoal do Nivel Superior (CAPES: AUX-PE-PROCAD 579/2008) and CNPq/PRONEX/FAPESQ.  A. Yu. P. has been supported by the CNPq project No. 303461-2009/8.

\end{document}